\documentclass[12pt]{article}
\usepackage{amsfonts}
\usepackage{amsmath}
\usepackage{amssymb}
\usepackage{color}
\usepackage{graphicx}
\usepackage{mathcomp}
\usepackage{setspace}
\usepackage[font=small,format=plain,labelfont=bf,up,textfont=it,up]{caption}

\makeatletter
\@addtoreset{equation}{section} %%% Latex Companion, p.~16
\makeatother

\def\appendix#1
{
\addtocounter{section}{1}\setcounter{equation}{0}%\setcounter{section}{1}
 \renewcommand{\thesection}{\Alph{section}}
 \section*{%Appendix~
 \thesection\protect\indent \parbox[t]{16.715cm}{#1}}
 \addcontentsline{toc}{section}{Appendix \thesection\ \ \ #1}
}
\begin{document}
\begin{center}
\textbf{Emergence of General Relativity from Loop Quantum Gravity: A Summary}
\end{center}
\begin{center}\textbf{Chun-Yen Lin}\\ \textit{Physics Department, University of California\\ Davis 95616, California}\\ 
\end{center}

{\small 
\begin{center}\textbf{Abstract}
\end{center}
A model is proposed to demonstrate that classical general relativity can emerge from loop quantum gravity, in a relational description of gravitational field in terms of the coordinates given by matter. Local Dirac observables and coherent states are defined to explore physical content of the model. Expectation values of commutators between the observables for the coherent states recover the four-dimensional diffeomorphism algebra and the large-scale dynamics of the gravitational field relative to the matter coordinates. Both results conform with general relativity up to calculable corrections near singularities.

\section{Introduction}

    Loop quantum gravity \cite{intro1,intro,perez} is a candidate quantum theory of gravity. Its non-perturbative approach strictly respect the background independence that is the heart of general relativity. Based on Dirac quantization in the Arnowitt-Deser-Misner (ADM) formalism \cite{intro1,intro,perez}, the theory's kinematic Hilbert space is rigorously defined and is called knot space. Knot space is spanned by knot states, where each knot state is a topological network colored by quantum numbers of gravitational and matter fields. The gravitational quantum numbers carried by the edges and nodes of the networks give quanta of area and volume in space \cite{intro1,intro,perez}. This quantum geometry successfully describes the discretized spatial geometry in Planck scales, and accounts for the Bekenstein-Hawking entropy of black holes \cite{gamma}.

Because of the background independence, the theory also faces new challenges \cite{intro1,intro,perez}: one has to \emph{solve} the intricate Hamiltonian constraint that acts on knot space to obtain a physical Hilbert space; and, one has to construct diffeomorphism-invariant local Dirac observables to describe the field dynamics. The difficulties make the semi-classical limit of the theory hard to obtain, and  it is unknown whether general relativity is a semi-clasical limit of loop quantum gravity. 

Following the guidance of loop quantum cosmology \cite{lqc1,lqc2,marolf1,marolf2}, the model in this paper obtains  its semi-classical limit from knot space by assuming: 1) a modified Hamiltonian constraint operator is valid; 2) a group averaging procedure applied to knot space solves the modified Hamiltonian constraint, resulting to the physical Hilbert space $\mathbb H$ for the model; 3) the matter back-reactions on the gravitational dynamics can be ignored in this context. The matter field operators provide spacetime internal coordinates for the gravitational local observables in the model \cite{torre1}\cite{torre}\cite{rovelli}. Appropriate coherent states in $\mathbb H$ are defined to minimize the uncertainty of the gravitational local observables, whose expectation values give rise to emergent classical gravitational fields. The symmetry of $\mathbb H$ then leads to equations governing the emergent gravitational fields. The equations reproduce classical general relativity in the vacuum, up to quantum gravitational corrections and matter back reactions. The corrections are expected to be important in small scales or near singular regions of the emergent spacetime, and the model also provides means to calculate them in the next step.

\section{Loop Quantum Gravity with Matter Fields}

\subsection{Knot space}
Loop quantum gravity is based on canonical general relativity in Ashtekar formalism \cite{form1,form2,form3}. This formalism describes gravitational fields in a form similar to that of matter gauge fields.

Traditionally, canonical general relativity uses spatial metric and extrinsic curvature defined in the spatial manifold $M$ as phase space variables \cite{form1,form2,form3}. As an alternative, triad fields consisting of three orthonormal vector fields $\{e^a_i (\text{x})\} (\text{x}\equiv (x,y,z); a=x,y,z;  i=1,2,3)$ can be used in place of the spatial metric. 
In the basis of $e^a_i (\text{x})$ and its inverse $e_a^i (\text{x})$, the spatial Levi-Civita connection and extrinsic curvature take the forms $\Gamma ^i_a (\text{x})$ and $K^i_a(\text{x})$. Ashtekar variables $(  A^i_a(\text{x}),E^a_i (\text{x}))$ are related to $(e^a_i (\text{x}), K^i_a(\text{x}))$ through a canonical transformation by \cite{form2,form3}:
\begin{equation}
\begin{split}
E^a_i (\text{x})\equiv {\det (e)} e^a_i(\text{x});\rule{10pt}{0pt}
A^i_a(\text{x}) \equiv \Gamma ^i_a (\text{x})+ \gamma K^i_a(\text{x})
\end{split}
\end{equation}
where the real number $\gamma$ is called the Immirzi parameter. By construction, the fields $E^a_i(\text{x})$ are densitized triad fields, and the fields $A^i_a(\text{x})$ are $SO(3)$ gauge fields. The variables have the non-vanishing Poisson brackets:
\begin{equation}
\begin{split}
\{A^i_a(\text{x}), E^b_j(\text{y})\}= 8\pi( G/c^3 )\gamma \delta^b_a \delta^i_j \delta (\text{x},\text{y})
\end{split}
\end{equation}
where $G$ is Newton's constant. Note that one may replace the $SO(3)$ symmetry group with $SU(2)$ in this formalism, since $SU(2)$ and $SO(3)$ share the same Lie algebra.

Matter fields with a gauge group $\mathcal G$ can be included in this formalism \cite{matter1,matter2,thiemann}. Using the triad basis we describe the fermion, scalar and gauge fields by $(\xi^{\bar i}_{\bar{\text i}}(\text{x}),\pi_{\bar i}^{\bar{\text i}}(\text{x}))$, $(\phi^{\text i}(\text{x}),P_{\text i}(\text{x}))$ and $(\underbar{A}^{\text i}_a(\text{x}),\underbar E_{\text i}^a(\text{x}))$. Here $\bar i$, ${\bar{\text i}}$ and ${\text i}$ are respectively (gravitational spin $ \frac{1}{2}$) $SU(2)$, $\mathcal G$ and adjoint $\mathcal G$ indices.

In this formalism, the action of general relativity possesses the following local symmetries: 1) time re-foliation invariance which introduces the Hamiltonian constraint $H(\bar N)=0$; 2) spatial diffeomorphism invariance, which introduces the momentum constraint $M(\bar V)=0$; 3) local $SU(2)$ invariance, which introduces the $SU(2)$ Gauss constraint $G(\bar{\Lambda})=0$; 4) local $\mathcal G$ invariance, which introduces the $\mathcal G$ Gauss constraint $\underbar{G}(\bar{\lambda})=0$. Each constraint is a functional of the dynamical fields and the Lagrangian multipliers $\bar N(\text{x})$, $\bar V^a(\text{x})$, $\bar{\Lambda}^i(\text{x})$, $\bar{\lambda}^{\text i}(\text{x})$ (the bars indicate their non-dynamical nature). The first three constraints result from spacetime diffeomorphism symmetry, and consist of pure gravitational and matter terms:
\begin{equation}
\begin{split}
H (\bar N)= H_g (\bar N)+ H_m (\bar N);\rule{10pt}{0pt}
G(\bar{\Lambda})= G_g(\bar{\Lambda})+ G_m(\bar{\Lambda});\rule{10pt}{0pt}
M(\bar V)= M_g(\bar V)+ M_m(\bar V)
\end{split}
\end{equation}
 Denoting $\kappa\equiv 8\pi( G/c^3 )$, the three constraints form a closed algebra with structure functionals independent of the explicit forms of the matter terms. Denoting $[\bar\Lambda, \bar\Lambda']$ to be the $SU(2)$ commutator, and setting $[ \bar V, \bar V']^a\equiv \bar V^b \partial_b  \bar V'^a- \bar V'^b \partial_b  \bar V^a$, the algebra is given by
\begin{equation}
\begin{split}
\{G(\bar{\Lambda}), G(\bar{\Lambda}')\}= \kappa\gamma G([\bar{\Lambda},\bar{\Lambda}']);\rule{10pt}{0pt}  \{G(\bar{\Lambda}), M(\bar V)\}= \kappa\gamma G(\mathcal{L}_{\bar V} \bar{\Lambda});\rule{10pt}{0pt}\{G(\bar{\Lambda}), H(\bar N)\}= 0\rule{20pt}{0pt} 
\\
\{M(\bar V), M(\bar V')\}=  \kappa \gamma M([\bar V,\bar V']);\rule{10pt}{0pt}
 \{M(\bar V), H(\bar N)\}=  \kappa \gamma H(\mathcal{L}_{\bar V}\bar N)\rule{70pt}{0pt}
\\
\{H(\bar N), H(\bar N')\}=\kappa\gamma\left( M(\bar S)+G(\bar S\cdot A)\right)+ {\kappa^{-1}}\gamma(1-\gamma^2) G\left(\frac{[E^a\partial_a \bar N,E^b\partial_b \bar N']}{|\det E|}\right)\rule{20pt}{0pt}
\end{split}
\end{equation}
where $\mathcal{L}_{\bar V}$ denotes a Lie derivative and
\begin{equation}
\begin{split}
\bar S^a=( \bar N\partial_b \bar N'-\bar N'\partial_b \bar N) \frac{E^b_i E^{ai}}{|\det E|}\rule{80pt}{0pt}\\
\end{split}
\end{equation}
 In QCD, the generalized electric flux and magnetic holonomy variables are powerful in capturing non-perturbative degrees of freedom of gluon fields. The Ashtekar formalism enables an analogous treatment of gravitational fields.
Gravitational holonomy and flux variables are signatures of loop quantum gravity \cite{intro,intro1,perez}, and they capture non-perturbative degrees of freedom of gravitational fields. The holonomy variable over an oriented path $\bar{e}\subset M$ (the bar indicates that $\bar{e}$ is embedded in the spatial manifold $M$) gives the parallel transport along the path by the connection fields $A^{i}_{b}(\text{x})$. The flux variable over an oriented surface $\bar{S}\subset M$ gives the flux of  ${E}^{a}_{i}(\text{x})$ through the surface. Explicitly, we have:
\begin{equation}
 {h}^{(j)}(\bar{e})^{\bar{k}}_{\bar{l}}[A]\equiv [\mathcal{P} \exp \int_{\bar{e}}d\bar{e}^{b} A^{i}_{b}(\text{x})\tau_{i}^{(j)}]^{\bar{k}}_{\bar{l}};\rule{10pt}{0pt}
 {F}_i(\bar{S})
\equiv \int_{\bar{S}}\hat{E}^{a}_{i}d\bar{S}_{a} 
\end{equation}
where $\mathcal{P}$ denotes path ordering along $\bar{e}$, and the $SU(2)$-valued gravitational holonomy is written in the spin $(j)$ matrix representation. The matter variables compatible with the gravitational flux and holonomy variables are the following  \cite{matter1,matter2,thiemann}. The gauge fields $\underbar{A}^{\text i}_a(\text{x})$ are described by the $\mathcal G$ holonomies $\text{h}^{(\text{i})}(\bar{e})^{\bar{\text i}}_{\bar{\text j}}$ in representations $(\text{i})$, and the $\underbar{E}_{\text i}^a(\text{x})$ fields are described by the flux variables ${\text F}_{\text i}(\bar{S})$. If $\bar v$ is a generic point in $M$, the $\xi^{\bar i}_{\bar{\text i}}(\text{x})$ fields are described by the irreducible tensors $\theta^{(d)}(\bar{v})$ obtained from their Grassmann monomials of degree $d$, and the spinor momenta $\pi_{\bar i}^{\bar{\text i}}(\text{x})$ are described by ${\eta}(\bar v)\equiv i{\theta}^{(d=1)}(\bar v)^{\dagger}$. The $\phi^{\text{j}}(\text{x})$ fields are described by $h^{(\text{k})}(\bar v) \equiv \exp(\phi^{\text{j}} (\bar v) {\tau^{(\text{k})}}_{\text{j}})$ called point holonomies in representations $(\text{k})$, and the momenta $P_{\text i}(\text{x})$  are described by ${p}_{\text i}(\bar v)$. This new set of gravitational and matter variables are collectively called \emph{loop variables}. The kinematical states of loop quantum gravity \cite{intro}\cite{intro1}\cite{perez} are called knot states, and they are functionals of the configuration Ashtekar variables $\{A^i_a(\text{x}),\underbar{A}^{\text i}_a(\text{x}),\xi^{\bar i}_{\bar{\text i}}(\text{x}),\phi _{\text i}(\text{x})\}$ via the corresponding loop variables $\{ {h}^{(j)}(\bar{e}),\text{h}^{(\text{i})}(\bar{e}), \theta^{(d)}(\bar{v}), h^{(\text{k})}(\bar v)\}$.

 A knot state is given by an $SU(2)\times \mathcal G$ invariant product of loop variables defined on a graph in $M$, so it solves the Gauss constraints. Further, the knot state is only sensitive to the embedding of the graph up to a spatial diffeomorphism $\mu\in \mathit{diff_M}$, so it also solves the momentum constraint. Define an embedded graph $\bar{\gamma}$ in $M$ to consist of $N_e$ smooth oriented paths $\{\bar{e}_i\}$, called edges, meeting at most at their end points $\{\bar{v}_n\}$, called nodes (the bars again indicate that $\bar{\gamma}$ is embedded in $M$). Carrying loop variables, an embedded colored graph $\bar \Gamma$ is defined by: 1) an embedded graph $\bar{\gamma}$; 2) an $SU(2)$ spin representation $j_{i}$ and a $\mathcal{G}$ group representation $\text{j}_i$ assigned to each edge;  3) generalized $SU(2)$ and $\mathcal{G}$ Clebsch-Gordan coefficients (intertwiners) $i_{n}$ and $\text{i}_n$, a point holonomy representation $\text{k}_n$ and a Grassmann monomial degree $d_n$ assigned to each node. The assignment of  $j_{i}$, $\text{j}_i$, $i_{n}$, $\text{i}_n$, $\text{k}_n$ and $d_n$ gives an $SU(2)\times \mathcal{G} $ scalar functional $S_{\bar{\Gamma}}$: 
\begin{equation}
\begin{split}
S_{\bar{\Gamma}}[A,\underbar{A}, \xi,\phi ]
\equiv Inv\left\{\bigotimes_n^{N_v} i_n \bigotimes_n^{N_v}\text{i}_n\bigotimes_n^{N_v}\theta^{(d_n)} (\bar{v}_n)\bigotimes_n^{N_v} h^{(\text{k}_n)}(\bar{v}_n) \bigotimes_i^{N_e}{h}^{(j_i)}(\bar{e}_i)  \bigotimes_i^{N_e}\text{h}^{(\text j_i)}(\bar{e}_i)\right\}[A,\underbar{A}, \xi,\phi]\\\\
\end{split}
\end{equation}
where $Inv\{ ...\}$ denotes the $SU(2) \otimes\mathcal{G}$ invariant contraction. 
 An element $\mu\in \mathit{diff_M}$ drags $\bar \Gamma$ to $\bar\Gamma'\equiv\mu\bar\Gamma$ and transforms $S_{\bar{\Gamma}}$ to $S_{\bar{\Gamma}}\hat\mu= S_{\bar\Gamma'}$. Note that for every $\bar\gamma$, there is a subgroup $\mathit{diff_{M,\bar{\gamma}}}$ of $\mathit{diff_M}$ that leaves the graph invariant, maintaining the \emph{set} of all the edges and their orientations. There is also a subgroup $\mathit{Tdiff_{M,\bar{\gamma}}}$ of $\mathit{diff_{M,\bar{\gamma}}}$ that acts trivially on $\bar \gamma$. Thus we have the graph symmetry group $G_{M,\bar{\gamma}}\equiv  \mathit{diff_{M,\bar{\gamma}}}/ \mathit{Tdiff_{M,\bar{\gamma}}}$. To erase the embedding information in $S_{\bar{\Gamma}}$, the group averaging operator $\hat{\mathbb P}_{\mathit{diff_M}}$ is defined as:
\begin{equation}
\begin{split}
S_{\bar{\Gamma}}\cdot\hat{\mathbb P}_{\mathit{diff_M}}\equiv S_{\bar{\Gamma}}\cdot\bigg[\frac{1}{N_{G_{M,\bar{\gamma}}}} \sum_{\mu \in G_{M,\bar{\gamma}}}\hat{\mu}\rule{3pt}{0pt} \cdot\rule{3pt}{0pt}\sum_{\mu'\in \mathit{diff_M}/\mathit{diff_{M,\bar{\gamma}}}}\hat\mu' \bigg]\equiv  s_{[\bar\Gamma]}
\end{split}
\end{equation}
where $N_{G_{M,\bar{\gamma}}}$ is the number of elements in $G_{M,\bar{\gamma}}$, and $[\bar\Gamma]$ is a colored graph obtained from the embedded colored graph $\bar\Gamma$ by erasing its exact embedding.\footnote{ Note that when $\bar\Gamma'=\bar\Gamma \hat\mu$, we have $[\bar\Gamma]=[\bar\Gamma']$.}
The result $s_{[\bar\Gamma]}$ is a knot state. Such a state is determined by a colored graph $[\bar\Gamma]$, and is $\mathit{diff_M}$ invariant since $s_{[\bar{\Gamma}]}\cdot \hat\mu=s_{[\mu\bar{\Gamma}]}= s_{[\bar{\Gamma}]}$.
The inner products between knot states are given by (generalized) Ashtekar-Lewandowski measure \cite{intro,intro1,perez,thiemann}, with which the set of all knot states $\{ \langle s_{[\bar\Gamma]}|\}$ gives an orthonormal basis:
\begin{equation}
\begin{split}
\langle s_{[\bar\Gamma']}|s_{[\bar\Gamma]}\rangle=\delta_{[\bar\Gamma'], [\bar\Gamma]}
\end{split}
\end{equation}
This basis spans knot space $K$, the $SU(2)\times \mathcal G$ and $\mathit{diff_M}$ invariant kinematic Hilbert space of loop quantum gravity.

Canonical quantization of loop variables leads to the operators of the form $\hat O(\bar{\Omega})$, where $\bar{\Omega}\subset M$ may be $\bar v$, $\bar e$ or $\bar S$. However, $\hat O(\bar{\Omega})$ does not preserve $K$ since $\bar{\Omega}$ is not $\mathit{diff_M}$ invariant.
To preserve $\mathit{diff_M}$ symmetry, we now replace $\bar{\Omega}$ by a dynamical object $\Omega$ that assigns $\bar\Omega( \bar\Gamma)\subset M$ to every $\bar\Gamma$, such that $\bar\Omega( \bar\Gamma)$ transforms together with $\bar\Gamma$ under $\mathit{diff_M}$ transformations. Setting $\{ \bar\Gamma_{rep}\}$ to contain one representative of every $[\bar\Gamma]$, one can construct $\Omega$ by specifying $\bar\Omega( \bar\Gamma_{rep})$ on the representatives $\{ \bar\Gamma_{rep}\}$. Formally, each dynamical object $\Omega$ is a map $\Omega: \bar\Gamma \to \bar{\Omega}(\bar\Gamma)\subset M$ satisfying $ \bar \Omega(\mu\bar \Gamma)= \mu'\mu \bar \Omega(\bar \Gamma)$ for any $\mu \in \mathit{diff_M}$ and some $\mu' \in  \mathit{Tdiff_{M,{\mu\bar\gamma}}}$. Then, we can define a $\mathit{diff_M}$ invariant operator $\hat O(\Omega) $ as:
\begin{equation}
\begin{split}
 s_{[\bar\Gamma]}\cdot \hat O(\Omega)\equiv S_{\bar\Gamma}\cdot\hat O(\bar{\Omega}(\bar\Gamma))\hat{\mathbb P}_{\mathit{diff_M}}\rule{10pt}{0pt}
\end{split}
\end{equation}
For example, one can construct $\mathit{diff_M}$ invariant flux and holonomy operators using a dynamical surface $S: \bar\Gamma \to \bar{S}(\bar\Gamma)\subset M$ and a dynamical path $e: \bar\Gamma \to \bar{e}(\bar\Gamma)\subset M$:
\begin{equation}
\begin{split}
 s_{[\bar\Gamma]}\cdot \hat{F}_i(S)\equiv S_{\bar\Gamma}\cdot\hat{F}_i(\bar{S}(\bar\Gamma))\hat{\mathbb P}_{\mathit{diff_M}};\rule{10pt}{0pt}
 s_{[\bar\Gamma]} \cdot \hat{h^{(j)}}({e})^{\bar{k}}_{\bar{l}}\equiv S_{\bar\Gamma}\cdot \hat{h^{(j)}}(\bar{e}(\bar\Gamma))^{\bar{k}}_{\bar{l}}\hat{\mathbb P}_{\mathit{diff_M}}
\end{split}
\end{equation}
The two operators are respectively differential and multiplicative operators acting on $s_{[\bar\Gamma]}$, and
their $SU(2)\times \mathcal G$ invariant products give gravitational operators in $K$.
Since $E^a_i (\text{x})$ determines the spatial metric, the spatial area and volume operators are made up of the flux operators \cite{intro,intro1,perez}. $\hat{F}_i(\bar S(\bar\Gamma))$ as a differential operator acts on the gravitational sector of $S_{\bar\Gamma}$ in a way that satisfies the Leibniz rule. Specifically, when $\bar S(\bar\Gamma)$ intersects with $\bar\gamma$ only at its node $\bar v_1$, we have:
\begin{equation}
\begin{split}
\hat{F}_{i}(\bar{S}(\bar\Gamma))\cdot \bigotimes_n^{N_v}  i_n \bigotimes_i^{N_e} {h}^{(j_i)}(\bar{e}_i)
\equiv\sum_{\bar e_{i'}|_{\bar v_1 \in \bar e_{i'}}} \iota(\bar{S},\bar{e}_{i'})\iota(\bar{e}_{i'}, \bar v_1)\hat{J}_{i}(\bar e_{i'}) \cdot \bigotimes_n^{N_v}  i_n \bigotimes_i^{N_e} {h}^{(j_i)}(\bar{e}_i)\rule{83pt}{0pt}\\
\hat{J}_{i}(\bar e)\cdot\bigotimes_n^{N_v}  i_n \bigotimes_i^{N_e} {h}^{(j_i)}(\bar{e}_i)
\equiv i\hbar\kappa\gamma \sum_k \left [\delta_{\bar e, \bar{e}_k}\cdot{h}^{(j_k)}(\bar{e}_k)\tau^{(j_k)}_i-\delta_{\bar e, {\bar{e}_k}^{-1}}\cdot\tau^{(j_k)}_i{h}^{(j_k)}(\bar{e}_k)\right]\bigotimes_{n}^{N_v}  i_n \bigotimes_{i\neq k}^{N_e} {h}^{(j_i)}(\bar{e}_i)\\\\
\end{split}
\end{equation}
where $\iota(\bar{S},\bar{e}_{i'})$ is $+1$ or $-1$ when $\bar{e}_{i'}$ is above or below $\bar S$ in an infinitesimal neighborhood of $\bar v_1$, and is $0$ if otherwise; $\iota(\bar{e}_{i'}, \bar v_1)$ is $+1$ or $-1$ when $\bar v_1$ is the source or target of $\bar{e}_{i'}$.

\section{The Model}

The remaining constraint to be imposed for a physical Hilbert space is the Hamiltonian constraint. Adhering to the polymer-like structure of knot states, the standard Hamiltonian constraint operator $\hat{H}(\bar N)^{LQG}\equiv\hat{H}^{LQG}_g(\bar N)+\hat{H}^{LQG}_m(\bar N)$ \cite{intro,perez,thiemann} is quantized from a regularized discrete expression approximating ${H}(\bar N)$. In the discrete expression, the curvature factors in ${H}(\bar N)$ are approximated by holonomies along a certain set of tiny loops. The quantization then leads to $\hat{H}(\bar N)^{LQG}$ that contains holonomy operators based on these loops. Therefore, the action of $\hat{H}(\bar N)^{LQG}$ on a knot state $s_{[\bar\Gamma]}$ involves a change in the graph topology that adds the set of tiny loops to $\bar\gamma$.
Moreover, since a non-constant Lagrangian multiplier $\bar N$ is not $\mathit{diff_M}$ invariant, $\hat{H}^{LQG}(\bar N)$ does not preserve $K$ in general. With a constant $\bar N$, the action of $\hat{H}^{LQG}(\bar N)$ preserves $K$, but it is intricate and changes the topology of the graphs. As a result, constructing a physical Hilbert space annihilated by $\hat{H}^{LQG}(\bar N)$ is a major challenge, which is currently tackled by both canonical approaches \cite{intro1,perez,group1,ave2} and the path integral formalism \cite{perez,foam2,foam1}. In our model, we will modify $\hat{H}^{LQG}(\bar N)$ into a graph topology preserving operator $\hat H( N_p)$ defined in $K$. The much simplified setting will allow us to apply group averaging method to construct the physical Hilbert space $\mathbb H$ of the model, based on certain concrete assumptions.

For each embedded graph $\bar\gamma$, we label each of its nodes with an integer $n$ and each of its edges connected to the node $n$ by an integer pair\footnote{Note that each edge has two labels since it contains two nodes, and the range of $i$ depends on $n$ in general} $(n,i)$. For a given $\bar\gamma$, we denote its node $n$ by $\bar v_n^{\bar\gamma}$, and the oriented path starting from $\bar v_n^{\bar\gamma}$ and overlapping exactly with its edge $(n,i)$ by $\bar e_{n,i}^{\bar\gamma}$ (fig.1a). To each pair $(\bar e_{n,i}^{\bar\gamma},\bar e_{n,j}^{\bar\gamma})$ we assign a minimal oriented closed path $\bar e_{n,i,j}^{\bar\gamma}$ that lies in $\bar\gamma$, containing the outgoing path $\bar e_{n,i}^{\bar\gamma}$ and incoming path $(\bar e_{n,j}^{\bar\gamma })^{-1}$ (fig.1a). Using these labels, we define a set of dynamical nodes $\{ v_m \}$, satisfying $\bar v_m(\bar\Gamma )=\bar v_{n(m)}^{\bar\gamma}$ such that $n(m)$ is one-to-one. Corresponding to a given $\{v_m\}$, we define a set of dynamical paths $\{ e_{m,j}\}$ satisfying $\bar e_{m,j}(\bar\Gamma )=\bar e_{n(m),k(j)}^{\bar\gamma }$ and $\bar v_m(\bar\Gamma )=\bar v_{n(m)}^{\bar\gamma}$ such that $j(k)$ is one-to-one. The dynamical closed path $e_{m,i,j}$ is then determined by the outgoing $ e_{m,i}$ and incoming $ e^{-1}_{m,j}$ dynamical paths. We also define a set of dynamical spatial points $p\equiv\{p_k\}$, where $k$ ranges from $0$ to infinity, such that $\bar v_n^{\bar\gamma}=\bar p_k(\bar\Gamma )$ holds for exactly one $k$ value for every $\bar\Gamma$ and $n$.\footnote{The spatial manifold $M$ contains uncountably many spatial points, but we use only countably infinite set $\{p_k\}$ in correspondence to the discrete structure of $K$. Also, the one-to-one correspondence between $\bar v_n^{\bar\gamma}$ and $\bar p_k(\bar\Gamma )$ is obvious had we taken $\{\bar p_k(\bar\Gamma)\}$ for any $\bar\Gamma$ to be the set of $all$ $points$ in $M$. Here we define $\{\bar p_k(\bar\Gamma)\}$ to be a countably infinite subset of $M$, while maintaining this natural condition.} Notice that there are infinitely many distinct sets of dynamical nodes, paths and spatial points satisfying the above. This ambiguity comes from the arbitrariness of identifying the nodes and edges between different knot states, due to the absence of a reference background in $K$.

\begin{figure}
\begin{center}
\includegraphics[angle=0,width=1.6in,clip=true]{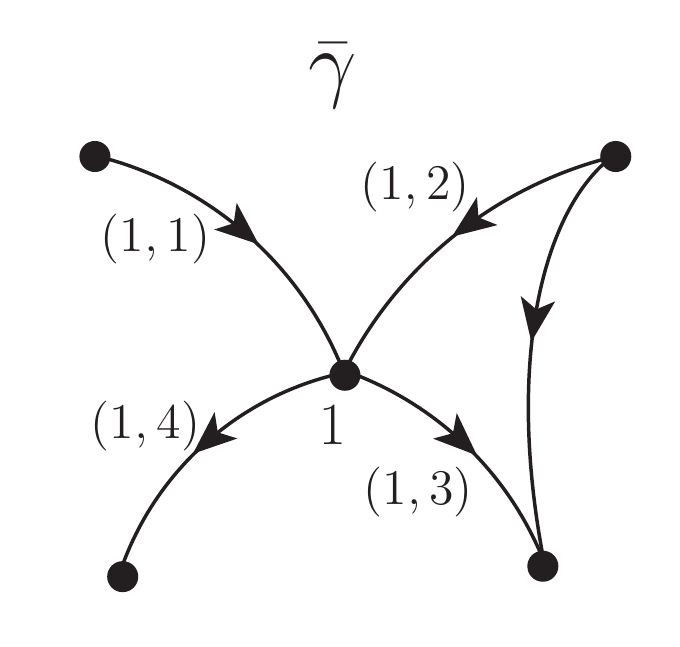}
\includegraphics[angle=0,width=4in,clip=true]{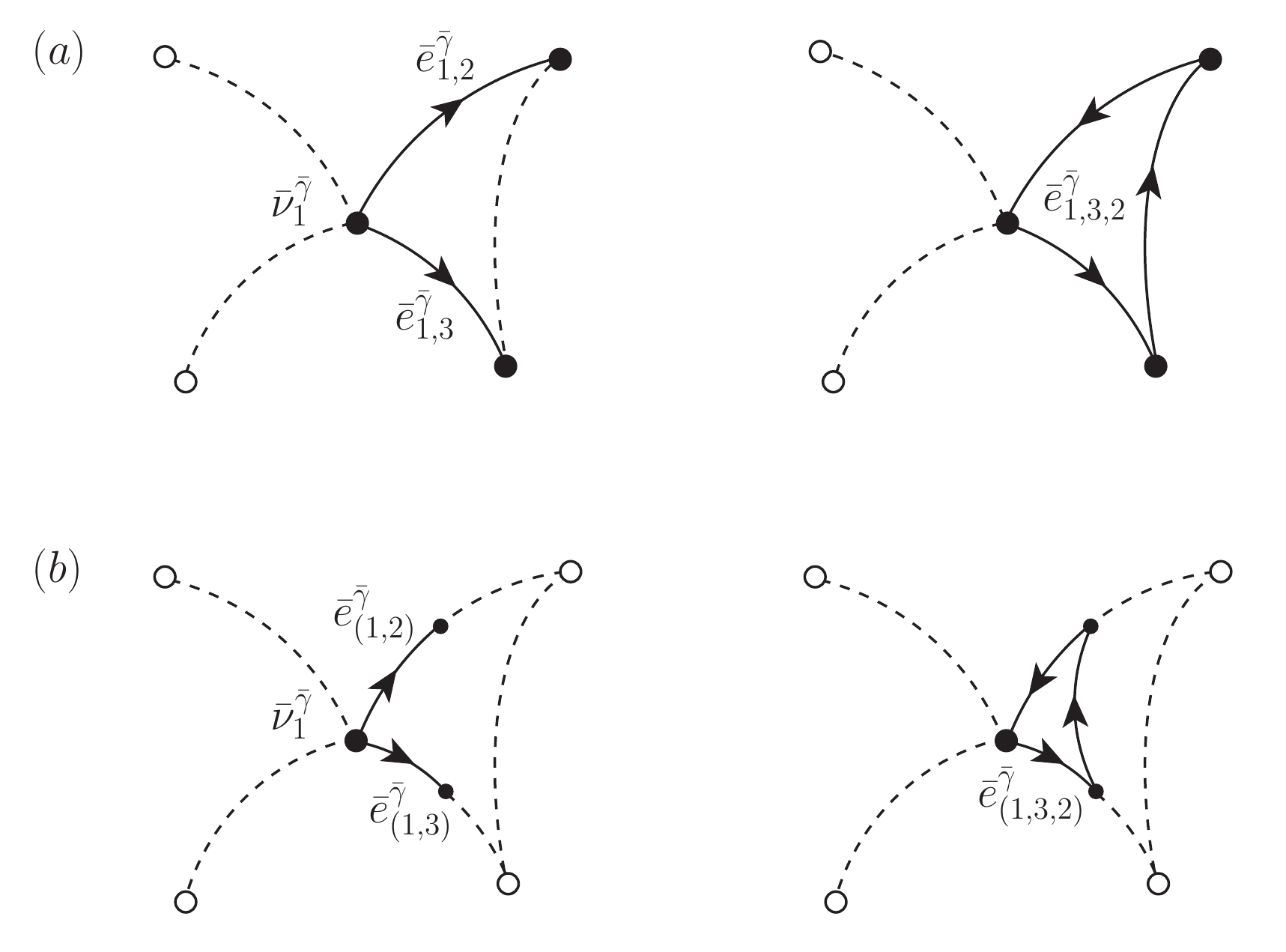}
\caption{The left figure depicts an embedded graph $\bar\gamma$, with one of its nodes labeled by $n=1$ and the four edges connected to this node labeled by $(1,j)$. The figures in $(a)$ demonstrate the corresponding definitions of $\bar v_n^{\bar\gamma}$, $\bar{e}^{\bar\gamma}_{n,j}$ and $\bar{e}^{\bar\gamma}_{n,i,j}$, while the figures in $(b)$ demonstrate the corresponding definitions of $\bar{e}^{\bar\gamma}_{(n,j)}$ and $\bar{e}^{\bar\gamma}_{(n,i,j)}$ that are used in $\hat{H}^{LQG}(\bar N)$.}
\end{center}
\end{figure}

The modification from $\hat{H}^{LQG}(\bar N)$ into $\hat H( N_p)$ -- a crucial step for the model -- contains two elements \cite{lin1}. First, $\bar N$ is replaced by a $\mathit{diff_M}$ invariant lapse function $N_p(p_k)$, which is a function of a set of dynamical spatial points $p=\{p_k\}$. Recall that the embedded spatial point $\bar p_k(\bar\Gamma)\in M$ transforms under $\mathit{diff_M}$ just as $\bar\Gamma$ does. That means the classical counterpart of $N_p(p_k)$ is the field $N(\text x)$ that transforms as a scalar field. Therefore, in proper semi-classical limits, the operator $\hat H( N_p)$ is expected to approximate the $\mathit{diff_M}$ invariant ${H}_{g}(N)$ instead of ${H}_{g}(\bar N)$. Second, the tiny loops (fig.1b) that define the holonomy operators in $\hat{H}^{LQG}(\bar N)$ are replaced by the new set of closed paths (fig.1a) that are contained in the graphs of  knot states. By using the corresponding new set of holonomy operators, $\hat H( N_p)$ preserves the graph-topology of knot states. Since $\hat H( N_p)$ preserves both $\mathit{diff_M}$ symmetry and the graph topology, it is an operator in any subspace $K_{\mathcal{T}}\subset K$ with one specific graph topology $\mathcal{T}$. For simplicity, the model's kinematical Hilbert space is set to be $K_{\mathcal{T}_{torus}}\subset K$ with the graph topology of a lattice torus $\mathcal{T}_{torus}$, which has $N_v$ nodes and six edges connected to each of the nodes. Restricted to $K_{\mathcal{T}_{torus}}$, we have $1\leq n\leq N_v$ and $1\leq i\leq 6$ for $v_n$ and $ e_{n,i}$, and $\bar e_{n,i,j}^{\bar\gamma}$ will be a square closed path overlapping with exactly four edges in $\bar\gamma$

The modification applied to the gravitational term of $\hat{H}^{LQG}(\bar N)$ results to the new gravitational term $\hat H'_g(N_p)$, which acts on $\langle s_{[\bar\Gamma]}|\in K_{\mathcal{T}_{torus}}$ as (set $\hat{h} (\bar e)\equiv\hat{h}^{(1/2)}(\bar e)$):
\begin{equation}
\begin{split}
 s_{[\bar\Gamma]}\hat{H}'_g(N_p)
\equiv S_{\bar\Gamma}\bigg[\hat{H}'^{E}_{g(\bar\Gamma)}(N_p) +\frac{4(1+\gamma^2)}{8\kappa^4\gamma^7(i\hbar)^{5}}\sum_{ v_m}  N_p(  p_k|_{ \bar p_k(\bar\Gamma )=\bar v_m(\bar\Gamma)}) \sum_{i,j,k=1} \text{sgn}\left(\bar e_{m,i}(\bar\Gamma ),\bar e_{m,j}(\bar\Gamma ),\bar e_{m,k}(\bar\Gamma )\right)\rule{110pt}{0pt}\\ \hat{h}^{-1} (\bar e_{m,i}(\bar\Gamma) )_{\bar{i}}^{\bar{l}} \left[\hat{h}(\bar e_{m,i}(\bar\Gamma ))_{\bar{l}}^{\bar{j}},\left[\hat{H}^{E}_{g(\bar\Gamma)}(1),\hat{V}_{(\bar\Gamma)}\right] \right]\hat{h}^{-1}(\bar e_{m,j}(\bar\Gamma ))_{\bar{j}}^{\bar{p}}\left[\hat h(\bar e_{m,j}(\bar\Gamma ))_{\bar{p}}^{\bar{k}},\left[\hat{H}^{E}_{g(\bar\Gamma)}(1),\hat{V}_{(\bar\Gamma)}\right] \right]\rule{100pt}{0pt}\\\times\hat{h}^{-1} (\bar e_{m,k}(\bar\Gamma )_{\bar{k}}^{\bar{q}}\left[\hat{h}(\bar e_{m,k}(\bar\Gamma )_{\bar{q}}^{\bar{i}},\hat{V}_{(\bar\Gamma)}\right]\bigg]\hat{\mathbb P}_{diff}\rule{335pt}{0pt}
\\
\end{split}
\end{equation}
 where we have
\begin{equation}
\begin{split}
\\
\hat{H}'^{E}_{g(\bar\Gamma)}(N_p)
\equiv \frac{2}{16\kappa^2\gamma(i\hbar)}\sum_{ v_m}   N_p(  p_k|_{ \bar p_k(\bar\Gamma )=\bar v_m(\bar\Gamma)})  \sum_{i,j,k=1} \text{sgn}\left(\bar e_{m,i}(\bar\Gamma ),\bar e_{m,j}(\bar\Gamma ),\bar e_{m,k}(\bar\Gamma )\right)\rule{200pt}{0pt}\\
      \left(\hat{ h}(\bar e_{m,i,j}(\bar\Gamma)) -\hat{ h}^{-1}(\bar e_{m,i,j}(\bar\Gamma))
      \right) ^{\bar{i}}_{\bar{j}} \left(\hat{h}^{-1} (\bar e_{m,k}(\bar\Gamma ))\right)_{\bar{i}}^{\bar{l}}\cdot \left[\left(\hat{h}(\bar e_{m,k}(\bar\Gamma ))\right)_{\bar{l}}^{\bar{j}},\hat{V}_{(\bar\Gamma)} \right]\rule{185pt}{0pt}
\end{split}
  \end{equation}
Here, the total volume operator $\hat{V}_{(\bar\Gamma)}$ is defined as
\begin{equation}
\begin{split}\\
\hat{V}_{(\bar\Gamma)}
\equiv \sum_{ v_m}\bigg[\frac{1}{48}\sum_{i,j,k=1} \text{sgn}\left(\bar e_{m,i}(\bar\Gamma ),\bar e_{m,j}(\bar\Gamma ),\bar e_{m,k}(\bar\Gamma )\right) \epsilon^{pqr}\hat{J}_{p}(\bar e_{m,i}(\bar\Gamma ))\hat{J}_{q}(\bar e_{m,j}(\bar\Gamma ))  \hat{J}_{r}(\bar e_{m,k}(\bar\Gamma ))\bigg]^{\frac{1}{2}}\rule{155pt}{0pt}\\
\end{split}
\end{equation}
The modification applied to the matter term in $\hat{H}^{LQG}(\bar N)$ \cite{thiemann} results to the new matter term $\hat{H}_m(N_p)$, whose explicit form depends on the matter content. Similar to the pure gravitational term, it is also constructed from the loop operators in $K_{\mathcal{T}_{torus}}$, and acts on $\langle s_{[\bar\Gamma]}|\in K_{\mathcal{T}_{torus}}$ as:
\begin{equation}
\begin{split}
  s_{[\bar\Gamma]}\hat{H}'_{m}(N_p)\equiv  S_{\bar\Gamma}\hat{H}'_{m(\bar\Gamma)}(N_p)\hat{\mathbb P}_{diff}\rule{350pt}{0pt}
\\
\hat{H}'_{m(\bar\Gamma)}(N_p)\equiv\sum_{ v_n} N_p(  p_k|_{ \bar p_k(\bar\Gamma )=\bar v_n(\bar\Gamma)})
\hat{H'}_{m}^{v_n}\bigg(\hat{J}_{i}(\bar e_{n,i}(\bar\Gamma )),\hat{h}(\bar e_{n,i}(\bar\Gamma ))_{\bar{l}}^{\bar{j}},\hat{h}( \bar e_{n,i,j}(\bar\Gamma))_{\bar{l}}^{\bar{j}},\hat{\text{J}}_{\text i}(\bar e_{n,i}(\bar\Gamma )),\rule{82pt}{0pt}\\ \hat{\text{h}}^{(\text j)}(\bar e_{n,i}(\bar\Gamma ))^{\bar{\text i}}_{\bar{\text j}}, \hat{\text{h}}^{(\text j)}(\bar e_{n,i,j}(\bar\Gamma))^{\bar{\text i}}_{\bar{\text j}}\hat {\theta} (\bar{ v}_n(\bar\Gamma ))^{\bar i}_{\bar{\text i}}, \hat {\theta} (\bar v_{n,i}(\bar\Gamma ))^{\bar i}_{\bar{\text i}},\hat{\eta} (\bar{ v}_n(\bar\Gamma ))_{\bar i}^{\bar{\text i}},\hat{\eta} (\bar v_{n,i}(\bar\Gamma ))_{\bar i}^{\bar{\text i}}, \hat{h}^{(\text{i})}(\bar{ v}_n(\bar\Gamma ))^{\bar{\text i}}_{\bar{\text j}},\rule{40pt}{0pt}\\\hat{h}^{(\text{i})}(\bar v_{n,i}(\bar\Gamma ))^{\bar{\text i}}_{\bar{\text j}},\hat{p}_{\text i}(\bar{ v}_m(\bar\Gamma )),\hat{p}_{\text i}(\bar v_{n,i}(\bar\Gamma )) \bigg)\rule{263pt}{0pt}\\
\end{split}
\end{equation}
where $v_{n,i}$ denotes the end node of $e_{n,i}$.

Finally, the Hamiltonian constraint operator for our model is given by the self-adjoint sum of the gravitational and matter terms: 
\begin{equation}
\hat H( N_p)\equiv \hat H_g(N_p)+\hat H_m(N_p)\equiv\frac{1}{2}(\hat H'_g(N_p)+\hat H'_m(N_p))+\frac{1}{2}(\hat H'_g(N_p)+\hat H'_m(N_p))^{\dagger}
\nonumber
\end{equation}

Recall that our kinematical Hilbert space $K_{\mathcal{T}_{torus}}$ is already $SU(2)\times \mathcal G$ and $\mathit{diff_M}$ invariant. To obtain the physical Hilbert space of the model, we still need to impose the remaining symmetry generated by $\{\exp(i\hat{H}(N_p))\}$ with arbitrary $N_p$ based on arbitrary $p$. These unitary operators form a faithful representation of a group $G$, that is
\begin{equation}
\left\{\prod_{k=1}^{\infty}\exp(i\hat{H}(N_{p_k}))\right\}_{\sim K_{\mathcal{T}_{torus}}}  \equiv \left\{ \hat{U}(g)\right\}_{g\in G}
\end{equation}
where $N_{p_k}$ is an arbitrary lapse function based on an arbitrary $p_k$, and $\sim K_{\mathcal{T}_{torus}}$ means that we identify two expressions if they give the same operator in $K_{\mathcal{T}_{torus}}$.
The physical Hilbert space of the model is constructed using group averaging procedure under the assumptions: (1) the existence of the left and right invariant measure $dg$ for $G$; (2) the operator $\hat{\mathbb{P}}$ defined by 
\begin{equation}
\hat{\mathbb{P}} \equiv \int_G dg \hat{U}(g) 
\end{equation}
maps $\langle \psi|\in K_{\mathcal{T}_{torus}}$ into $\hat{\mathbb P}|\psi\rangle\in K_{\mathcal{T}_{torus}}^*$. The two conditions hold in minisuperspace models \cite{ave1,marolf1,marolf2}, but remain to be proven for the model.
Under these assumptions, the inner product between any two states $|\Psi_1\rangle=\hat{\mathbb{P}}|\psi_1\rangle$ and $| \Psi_2\rangle=\hat{\mathbb{P}}|\psi_2\rangle$ may be defined as
\begin{equation}
\langle\Psi_1|\Psi_2\rangle\equiv\langle\psi_1|\Psi_2\rangle= \langle\psi_1|\hat{\mathbb{P}}|\psi_2\rangle \\
\end{equation}

The physical Hilbert space of the model $\mathbb H \subset  K_{\mathcal{T}_{torus}}^*$ is the space spanned by $\{\hat{\mathbb P}|s_{[\bar\Gamma]}\rangle\}$. By construction, $\mathbb H$ is invariant under the action of $\{\exp(i\hat{H}(N_p))\}$ with arbitrary $N_{p}$, and satisfies the modified Hamiltonian constraint. Each element in $\mathbb H$ is a solution to the quantized  Einstein equations, and therefore is a quantum state of spacetime with the matter fields.

\subsection{Local Dirac Observables}

The model obtains its local Dirac observables using clocks, spatial coordinates and frames given by the matter fields. Given any set of dynamical nodes $\{v_m\}$, one may build a set of self-adjoint and commuting matter operators \cite{lin1} consisting of scalar operators $\{ \hat\phi^{0}(v_m),\hat\phi^{1}(v_m), \hat\phi^{2}(v_m), \hat\phi^{3}(v_m)\}$ diagonalized by the knot state basis of $K_{\mathcal{T}_{torus}}$, current operators $\{\hat{V^{i}_{I}}(v_m), \hat{U}^{\bar{i}}_{\bar{I}}(v_m)\}$ and conjugate current operators $\{\hat{\bar{V}}^{I}_{i}(v_m),\hat{\bar{U}}^{\bar{I}}_{\bar{i}} (v_m)\}$ ($I=1,2, 3$ for the vector currents; $\bar{I}=1,2$ for the spinor currents). The operators $(\hat{\phi}^{1}(v_m), \hat{\phi}^{2}(v_m), \hat{\phi}^{3}(v_m))\equiv \hat{\Phi} (v_m)$ will serve as spatial coordinate operators, $\{\hat{V^{i}_{I}}(v_m), \hat{U}^{\bar{i}}_{\bar{I}}(v_m)\}$ and $\{\hat{\bar{V}}^{I}_{i}(v_m), \hat{\bar{U}}^{\bar{I}}_{\bar{i}} (v_m)\}$ will serve as spatial frame operators, and $\hat\phi^{0}(v_m)$ will be the clock operator. 

Classical gravitational fields in the Ashtekar formalism are $SU(2)$ tensors. In the model, the $SU(2)$ invariant components of the fields are described relative to the matter spatial frames. Explicitly, for any $ s_{[\bar\Gamma]}\in K_{\mathcal{T}_{torus}}$ we define
\begin{equation}
\begin{split}
s_{[\bar\Gamma]}\cdot\hat{J}( e_{n, j})_{I} \equiv S_{\bar\Gamma} \cdot\hat{V}(\bar v_n(\bar \Gamma))^{i}_{I}\hat{J}(\bar e_{n, j}(\bar \Gamma))_{i}\hat{\mathbb P}_{diff}\rule{55pt}{0pt}\\
 s_{[\bar\Gamma]}\cdot\hat{h}(e_{n,k})^{\bar{I}}_{\bar{J}} \equiv  S_{\bar\Gamma}\cdot\hat{\bar{U}}(\bar v_{n,k}(\bar \Gamma))^{\bar{I}}_{\bar{i}} \hat{h}(\bar e_{n,k}(\bar \Gamma))^{\bar{i}}_{\bar{j}}\hat{U}(\bar v_n(\bar \Gamma))^{\bar{j}}_{\bar{J}}\hat{\mathbb P}_{diff}
\end{split}
\end{equation}
In canonical general relativity with the matter spatial coordinate field $\Phi(\text x)$, we can obtain a $\mathit{diff_M}$ invariant and spatially local variable $O(X)$ by integrating $\det(\partial\Phi(\text x))\delta(\Phi(\text x)-X) O(\text x)$ over $M$. Analogously, the model uses a normalized Gaussian distribution ${\delta}^{\epsilon}$ with finite width $\epsilon$ and defines:
\begin{equation}
\begin{split}
s_{[\bar\Gamma]}\hat{O}(X) \equiv  S_{\bar\Gamma}\sum_{n}\det(\Delta \hat{\Phi}(\bar v_n(\bar\Gamma))) \hat{\delta}^{\epsilon}( \hat{\Phi}(\bar v_n(\bar\Gamma))- X)\hat{O}(\bar v_n(\bar\Gamma))  \hat{\mathbb P}_{diff}\rule{135pt}{0pt}\\
s_{[\bar\Gamma]}\hat{O'}(e_{X,\Delta X})\equiv S_{\bar\Gamma}\sum_{n,i} \det(\Delta \hat{\Phi}(\bar v_n(\bar\Gamma))) \det(\Delta \hat{\Phi}(\bar v_{n,i}(\bar\Gamma)))\hat{\delta}^{\epsilon}(\hat{\Phi}(\bar v_n(\bar\Gamma))-X) \hat{\delta}^{\epsilon}(\hat{\Phi}(\bar v_{n,i}(\bar\Gamma))-X-\Delta X)\\\times\hat{O}(\bar e_{n,i}(\bar\Gamma))\hat{\mathbb P}_{diff}\rule{315pt}{0pt}\\
\end{split}
\end{equation}
 for any $ s_{[\bar\Gamma]}\in K_{\mathcal{T}_{torus}}$.
where the coordinate volume element operators are given by:
\begin{equation}
\begin{split}
\Delta \hat{\Phi}_{\bar e_{n,i}(\bar \Gamma)}\equiv  \big[\hat{\Phi}(\bar v_{n,i}(\bar\Gamma)) - \hat{\Phi}(\bar v_n(\bar \Gamma))\big]\rule{275pt}{0pt}
\\
 S_{\bar\Gamma}\det(\Delta \hat{\Phi}(\bar v_n(\bar \Gamma)))
\equiv  S_{\bar\Gamma}\sum_{(i,j,k)}\text{sgn}\left(\bar e_{n,i}(\bar \Gamma),\bar e_{n,j}(\bar \Gamma),\bar e_{n,k}(\bar \Gamma)\right) \Delta \hat{\Phi}_{\bar e_{n,i}(\bar \Gamma)}\cdot \left(\Delta \hat{\Phi}_{\bar e_{n,j}(\bar \Gamma)} \times\Delta \hat{\Phi}_{\bar e_{n,k}(\bar \Gamma)}\right)
\nonumber
\end{split}
\end{equation}
 It is crucial that the spatially local operators obtained in $(3.9)$ do not depend on the choices of the dummy variables $\{v_m\}$ and $\{e_{n,i}\}$. The $\mathit{diff_M}$ invariant operators are localized by referring to the spatial matter coordinates.
 
To specify time using the clock field $\phi^{0}$, the model constructs an operator $\hat{\Pi}_T$ that maps a spacetime state $|\Psi\rangle\in \mathbb H$ to its spatial slice state $\hat{\Pi}_T|\Psi\rangle\in K^*_{\mathcal{T}_{torus}}$ where $\phi^{0}=T$. With $\hat{\omega}(\bar v)$ denoting an operator acting on $\bar v$, the operator $\hat{\Pi}_T$ is defined by:
\begin{equation}
\begin{split}
\hat{\nu}_{\omega} (\bar v_n(\bar\Gamma)) \equiv \frac{ i}{\hbar}\left[\hat{\omega}(\bar v_n(\bar\Gamma)), \hat{H}_{(\bar \Gamma)} (1) \right]\rule{111pt}{0pt}
\\
 s_{[\bar\Gamma]}\hat{\Pi}_T\equiv S_{\bar\Gamma} sym\left\{\prod_n \hat{\nu}_{\phi^0} (\bar v_n(\bar\Gamma)) \hat{\delta}^{\epsilon}(\hat{\phi}^0(\bar v_n(\bar\Gamma))- T)\right\}\hat{\mathbb P}_{diff}\rule{3pt}{0pt}
\end{split}
\end{equation}
where the symmetrization $sym\{...\}$  in the ordering of $n$ is applied. The resulting local Dirac observables in $\mathbb H$ are given by:
\begin{equation}
\hat{O}(X, T)\equiv\hat{\mathbb{P}}\hat{O}(X)  \hat{\Pi}_T\rule{2pt}{0pt};\rule{4pt}{0pt}\hat{O'}(e_{X,\Delta X},T)\equiv\hat{\mathbb{P}}\hat{O'}(e_{X,\Delta X})  \hat{\Pi}_T
\end{equation}
In the following, $\omega$ in $(3.10)$ will serve as an index running over the matter coordinates and frames $\{\phi^0,\rule{2pt}{0pt}\Phi, \rule{3pt}{0pt}f\cdot V_I\rule{1pt}{0pt},\rule{3pt}{0pt} \bar f\cdot\bar V^I\rule{0pt}{0pt} ,\rule{3pt}{0pt}g\cdot U_{\bar I}\rule{1pt}{0pt}, \rule{3pt}{0pt}\bar g \cdot\bar U^{\bar I}\}$ contracted with the non-zero $SU(2)$ test functions $\{f^i,\bar f_i, g^{\bar i}, \bar g_{\bar i}\}$. The variables ${\nu}_{\omega}$ thus contain information about the momenta of the reference matter fields.

Finally, applying $(3.11)$ to the fields of concern we obtain the relevant localized observables $\hat{J}(e_{X, \Delta X}, T)_{I}$, $\hat{h}(e_{X, \Delta X}, T)^{\bar{I}}_{\bar{J}}$ and $\hat{\nu}_{\omega}(X, T)$.

\subsection{Conditions on Matter Coordinates and Frames}

Next, we choose a proper physical state $|\Psi\rangle \in \mathbb H$ to derive semi-classical limits from the model.
 In order to give sensible descriptions, the localized observables must refer to well-behaved matter coordinates and frames. So there are conditions to be imposed on the matter sector of $|\Psi\rangle$, which provides the matter coordinates and frames. 

For the clock, we require that any two spatial slices of $|\Psi\rangle$ with $\phi^{0}=T_1$ and $\phi^{0}=T_2$ are related by causal dynamics, and thus either one of them can be used to reconstruct the spacetime. In the model, this condition is imposed as:
\begin{equation}
\hat{ \mathbb{P}}\hat{\Pi}_{T_1}|\Psi\rangle+O(\hbar) = \hat{\mathbb{P}}\hat{\Pi}_{T_2}|\Psi\rangle+O(\hbar)=|\Psi\rangle
\end{equation}

On each spatial slice $\hat{\Pi}_{T_1}|\Psi\rangle$, we will also impose spatial coordinate conditions.
 Denoting an arbitrary combination of gravitational operators based on $\{ e_{n,i}\}$ as $ \varphi( \hat{J}( e_{n,i})_{i} ,\hat{h} (e_{n,i})^{\bar{i}}_{\bar{j}})$, we require that there exist $\{v^*_n\}$ and $\{e^*_{n,i}\}$ such that:
\begin{equation}
\begin{split}
\varphi( \hat{J}(e^*_{n,i})_{i} ,\hat{h}(e^*_{n,i})^{\bar{i}}_{\bar{j}},\hat{h}^{\dagger}(e^*_{n,i})^{\bar{i}}_{\bar{j}})\hat{\Pi}_{T_1}|\Psi\rangle\rule{125pt}{0pt}\\
= \varphi( \hat{J}(e_{X_n, \Delta X_{n,i}})_{I} ,\hat{h}(e_{X_n, \Delta X_{n,i}})^{\bar{I}}_{\bar{J}},\hat{h}^{\dagger}(e_{X_n, \Delta X_{n,i}})^{\bar{I}}_{\bar{J}})\hat{\Pi}_{T_1}|\Psi\rangle+O(\hbar)
\end{split}
\end{equation}
with a certain set of values $\{X_n\}$ and $\{\Delta X_{m,j}\}$ satisfying $\{X_m+\Delta X_{m,j}\}=\{X_n\}$. Here, $\{v^*_ n\}$ represents the physical nodes at clock time $T_1$ for an observer using the spatial matter coordinates $\Phi$. Naturally, there is also a set of physical spatial points $p^*\equiv \{p^*_m\}$ agreeing with $\{v^*_n\}$, such that $ \bar p^*_m(\bar\Gamma)=\bar v^*_m(\bar\Gamma)$ for $m \le N_v$. Condition $(3.13)$ says that each physical node acquires a matter spatial coordinate value, and that the matter frames are physically orthonormal to each other at each physical node.
The spatial coordinate condition can now be imposed on the map $( v^*_n, e^*_{n,i}) \to (X_n, \Delta X_{n,i})$. Analogous to the coordinate maps on a torus manifold, the map should appear smooth in large scales at most of the physical nodes except at some $\{v^*_{n_b} \}\subset \{ v^*_{n}\}$ where coordinate singularities occur. We demand $|\Delta X_{n,i}| \leq d$ for $v^*_n \not\in \{ v^*_{n_b}\}$, where $d$ is small enough that the map appears continuous at large scales. Also, for any $m$ the set $\{\Delta X_{m,i}\leq d\}$ should define a parallelepiped in $\mathbb{R}^3$ up to an error of $O(d)$ so the map appears smooth at large scales. Notice that once $(3.13)$ is satisfied, the coordinate conditions can be achieved easily through a redefinition of coordinates  $\hat\Phi \to \hat\Phi'(\hat\Phi)$.

Using a lapse function $ N^{\small \mathcal{N}}_{p^*}$ satisfying $N^{\small \mathcal N}_{p^*}(p^*_m) = \mathcal{N}(X_{m})$ with an arbitrary function $\mathcal{N}(X)$, we can apply $(3.13)$ to $\hat{H}_{g}( N^{\small \mathcal{N}}_{p^*})$:
\begin{equation}
\begin{split}
\hat{H}_{g}( N^{\small \mathcal{N}}_{p^*})|\Psi\rangle
\equiv\varphi( N^{\small \mathcal{N}}_{p^*}(p^*_n),\hat{J}( e^*_{n,i})_{i} ,\hat{h} (e^*_{n,i})^{\bar{i}}_{\bar{j}})|\Psi\rangle
= \hat{\mathcal{H}}_g(\mathcal N)\hat{\Pi}_{T_1}|\Psi\rangle+O(\hbar)\\
\hat{\mathcal{H}}_g(\mathcal N)\equiv \varphi(\mathcal{N}(X_n),\hat{J}(e_{X_n, \Delta X_{n,i}})_{I} ,\hat{h}(e_{X_n, \Delta X_{n,i}})^{\bar{I}}_{\bar{J}})  \rule{79pt}{0pt}
\end{split}
\end{equation}
 Finally, the well-behaved matter coordinates and frames also lead to the algebraic relations between the spatially local operators, when they act on $\mathit{to}$ $\varphi( \hat{J}( e^*_{n,i})_{i} ,\hat{h} (e^*_{n,i})^{\bar{i}}_{\bar{j}})\hat{\Pi}_{T_1}|\Psi\rangle\equiv |\varphi\rangle$. From now on we set $(X, \Delta X)\in \{(X_n, \Delta X_{n,i})\}$.  Also, define $\hat{\alpha}(e_{Y, \Delta Y})$ as the operator localized from $\hat{\alpha}(e_{n,i})$ which is diagonalized by the basis $\{\langle s_{[\bar \Gamma]}| \in K_{\mathcal{T}_{torus}}\}$. The state $\langle s_{[\bar \Gamma]}|$ has eigenvalue $+1$ if the embedded edge of $\bar \Gamma$ overlapping with $\bar e_{n,i}(\bar \Gamma)$ has the same orientation as $\bar e_{n,i}(\bar \Gamma)$ and has eigenvalue $-1$ if otherwise. The algebraic relations are:
\begin{equation}
\begin{split}
\hat{J}^\dagger(e_{X, \Delta X})_{I}|\varphi\rangle=\hat{J}(e_{X, \Delta X})_{I} |\varphi\rangle+O(\hbar)\rule{153pt}{0pt}\\
\left[\hat{J}(e_{X, \Delta X})_{I}, \hat{h}(e_{Y, \Delta Y})^{\bar{I}}_{\bar{J}}\right]|\varphi\rangle= \delta_{X,Y} \delta_{\Delta X,\Delta Y}i l_p^2\gamma  ({\tau _I})^{\bar{K}}_{\bar{J}} \hat{h}(e_{Y, \Delta Y})^{\bar{I}}_{\bar{K}}|\varphi\rangle\rule{87pt}{0pt}\\ - \delta_{X,Y+ \Delta Y} \delta_{-\Delta X,\Delta Y}i l_p^2 \gamma({\tau _I})^{\bar{I}}_{\bar{K}} \hat{h}(e_{Y, \Delta Y})^{\bar{K}}_{\bar{J}}|\varphi\rangle+O( l_p^2 \hbar)\\
\left[\hat{J}(e_{X, \Delta X})_{I}, \hat{h}^{\dagger}(e_{Y, \Delta Y})^{\bar{I}}_{\bar{J}}\right]|\varphi\rangle= \delta_{X,Y} \delta_{\Delta X,\Delta Y}i l_p^2 \gamma ({\tau^* _I})^{\bar{K}}_{\bar{J}} \hat{h}^{\dagger}(e_{Y, \Delta Y})^{\bar{I}}_{\bar{K}}|\varphi\rangle\rule{79pt}{0pt}\\ - \delta_{X,Y+ \Delta Y} \delta_{-\Delta X,\Delta Y}i l_p^2\gamma ({\tau^*_I})^{\bar{I}}_{\bar{K}} \hat{h}^{\dagger}(e_{Y, \Delta Y})^{\bar{K}}_{\bar{J}}|\varphi\rangle+O( l_p^2 \hbar)\\
\left[\hat{J}(e_{X, \Delta X})_{I}, \hat{J}(e_{Y, \Delta Y})_{J}\right]|\varphi\rangle= \delta_{X,Y} \delta_{\Delta X,\Delta Y}i l_p^2\gamma{\epsilon_{IJ}}^K \hat{J}(e_{Y, \Delta Y})_{K} \hat{\alpha}(e_{Y, \Delta Y})|\varphi\rangle+O( l_p^2  \hbar)\\
\end{split}
\end{equation}
These relations enable further calculations after the approximation $(3.13)$ is made.

\subsection{ Coherent States and Emergent Fields }

Suppose $|\Psi\rangle$ satisfies the matter coordinate and frame conditions $(3.12)$ and $(3.13)$, such that the local observables $\{\hat{J}(e_{X, \Delta X}, T)_{I},\hat{h}(e_{X, \Delta X}, T)^{\bar{I}}_{\bar{J}}, \hat{\nu}_{\omega}(X, T_{1})\}$ give meaningful descriptions around the moment $T_1$. Since our goal is to obtain semi-classical limits, we impose coherence conditions on $|\Psi\rangle$ with respect to the observables:
\begin{equation}
\begin{split}
\hat{J}(e_{X, \Delta X}, T_1)_{I}|\Psi\rangle=\langle\hat{J}(e_{X, \Delta X}, T_1)_{I}\rangle |\Psi\rangle +O(l_p^2) ;\rule{2pt}{0pt}
\hat{h}(e_{X, \Delta X}, T_1)^{\bar{I}}_{\bar{J}}|\Psi\rangle=\langle\hat{h}(e_{X, \Delta X}, T_1)^{\bar{I}}_{\bar{J}}\rangle |\Psi\rangle +O(l_p^2)\\
\hat{\nu}_{\omega}(X, T_{1})|\Psi\rangle=\langle\hat{\nu}_{\omega}(X, T_{1})\rangle|\Psi\rangle +O(\hbar)\rule{275pt}{0pt}
\end{split}
\end{equation} 
such that for any two $\bar e^*_{n,i}(\bar\Gamma)$ and $\bar e^*_{m,j}(\bar\Gamma)$ that share a common node and form a smooth path, the expectation values satisfy (recall that $|\Delta X_{n,i}| \leq d$ for $v^*_n \not\in \{ v^*_{n_b}\}$):
\begin{equation}
\begin{split}
\langle\hat{J}(e_{X_n, \Delta X_{n,i}}, T_1)_{I}\rangle= \langle\hat{J}(e_{X_m, \Delta X_{m,j}}, T_1)_{I}\rangle +O(d);\rule{1pt}{0pt}
\langle\hat{h}(e_{X_n, \Delta X_{n,i}}, T_1)^{\bar{I}}_{\bar{J}}\rangle=\langle\hat{h}(e_{X_m, \Delta X_{m,j}}, T_1)^{\bar{I}}_{\bar{J}}\rangle+O(d)\\
\langle\hat{\nu}_{\omega}(X_n, T_{1})\rangle=\langle\hat{\nu}_{\omega}(X_n+\Delta X_{n,i}, T_{1})\rangle|\Psi\rangle +O(d)\rule{237pt}{0pt}
\end{split}
\end{equation} 
Because of the algebraic relations in $(3.15)$, we expect the solutions to $(3.16)$ and $(3.17)$ to exist. The explicit construction of flux coherent states in loop quantum gravity is a subtle issue, and it is rigorously studied in works such as \cite{chr1,chr2}. In the following, we will involve only the semi-classical properties $(3.16)$ and $(3.17)$ of $|\Psi\rangle$, which imply that the quantum spacetime has sharply defined and approximately continuous values for the local observables at the clock time $T_1$. 

To make contact with classical general relativity, the model maps the expectation values in $(3.16)$ to the classical field values through an algorithm, using the matter coordinates as a common reference. Suppose $\{X_n\}$ approximatedly occupies a region $\bar I^3\subset \mathbb R^3$. Then each physical dynamical path $ e^*_{n.i}$ is identified with an oriented smooth path $\bar{e}_{X_n, \Delta X_{n,i}}\subset \bar I^3 $ going from $X_n$ to $X_n+\Delta X_{n,i}$. Also,  $\bar{S}_{X_n, \Delta X_{n,i}}\subset\bar I^3$ is defined to be a rectangular, oriented surface that intersects $\bar{e}_{X_n, \Delta X_{n,i}}$ with the same orientation, such that $\{\bar{S}_{X_n, \Delta X_{n,i}}\}$ defines a cell decomposition of $\bar I^3 $ that is dual to the lattice defined by $\{\bar{e}_{X_n, \Delta X_{n,i}}\}$. After the identification, the algorithm maps the expectation values $\{\langle \hat{J}(e_{X, \Delta X}, T_1)_{I}\rangle, \langle\hat{h}(e_{X, \Delta X}, T_1)^{\bar{I}}_{\bar{J}}\rangle,\langle\hat{\nu}_{\omega}(X, T_1)\rangle\}$ to the smooth fields $\{ E^{a}_{I}( X, T_1 ), A^{J}_{b}( X, T_1),{\nu}_{\omega}(X, T_1)\}$ defined in $\bar I^{3}$. The map obeys the following rules:\footnote{ Note that such a map is guaranteed to exist, since we are fitting the smooth fields with infinitely many degrees of freedom to the finitely many data points given by the expectation values of the local observables.}
\begin{equation}
\begin{split}
  \int_{\bar{S}_{X, \Delta X}}E^{a}_{I}(T_1)ds_{a}  \equiv \langle \widehat{J}(e_{X, \Delta X}, T_1 )_I \rangle+O(l_p^2)\rule{180pt}{0pt}\\
\mathcal{P}\exp[ \int_{\bar{e}_{X, \Delta X}} A^{J}_{b}(T_1)(\tau_{J})de^{b}]^{\bar{K}}_{\bar{L}} \equiv \langle \hat{h}(e_{X, \Delta X}, T_1)^{\bar{K}}_{\bar{L}}\rangle +O(l_p^2);\rule{3pt}{0pt}
{\nu}_{\omega}(X, T_1)\equiv\langle\hat{\nu}_{\omega}(X, T_1)\rangle+O(\hbar)\rule{0pt}{0pt}
\end{split}
\end{equation}
The fitting algorithm described above is restricted but non-unique. However, any fit satisfying the requirements $(3.18)$ gives a valid correspondence between $|\Psi\rangle$ and the smooth fields at $T_1$.

\subsection{Semi-Classical Limit of $\Psi$}
Sequentially applying the conditions $(3.15)$, $(3.12)$, $(3.16)$, and $(3.18)$, one finds that (for details see \cite{lin1}):
\begin{equation}
\begin{split}
\langle\Psi|\hat{\mathbb P}\left(\frac{i}{\hbar}\right)^{n-1}\left[\hat{\mathcal H}_{g}(\mathcal{N}_n),....\left[\hat{\mathcal H}_{g}(\mathcal{N}_3),\left[\hat{\mathcal H}_{g}(\mathcal{N}_2), \hat{\mathcal H}_{g}(\mathcal{N}_1)\right]\right]...\right]\hat{\Pi}_{T_1}|\Psi\rangle\rule{80pt}{0pt}\\
=\left\{{H}_{g}(\bar{N}_n),....\left\{{H}_{g}(\bar N_3),\left\{{H}_{g}(\bar{N}_2),{H}_{g}(\bar{N}_1)\right\}\right\}...\right\}\big|_{E_{I}^{a}(T_1), A_{b}^{J}(T_1), \bar{N}_i=\mathcal{N}_i}
+O(\hbar)+O(d^4)\\
\end{split}
\end{equation}
with arbitrary $n$ and $\mathcal{N}_i$. According to $(2.4)$ in the case of an empty matter sector, $(3.19)$ reproduces the full (off-shell ) algebra between $ H_{g}(\bar N)$, $G_{g}(\bar{\Lambda})$ and $M_{g}(\bar V)$ in the semi-classical limit of $|\Psi\rangle$, up to the corrections $O(\hbar)+O(d^4)$. 

Moreover, the symmetry $\hat{H}(N_{p})|\Psi\rangle=0$ has two major implications when we choose $N_{p}=N^{\mathcal{N}}_{p^*}$ (for details see \cite{lin1}):

(1) With arbitrary $n$ and $\mathcal{N}_i$, the symmetry implies that:
\begin{equation}
\begin{split}
 \langle\Psi|\hat{\mathbb P}\left(\frac{i}{\hbar}\right)^{n-1}\left[\hat{H}(N^{\mathcal{N}_n}_{p^*}),....\left[\hat{H}(N^{\mathcal{N}_3}_{p^*}),\left[\hat{H}(N^{\mathcal{N}_2}_{p^*}), \hat{H}(N^{\mathcal{N}_1}_{p^*})\right]\right]...\right]\hat{\Pi}_{T_1}|\Psi\rangle=0
\end{split}
\end{equation}
One can evaluate the purely gravitational contribution to $(3.20)$ involving only $\hat{H}_g(N^{\mathcal{N}_i}_{p^*})$, which can be approximated  by $\hat{\mathcal H}_{g}(\mathcal{N}_i)$ according to $(3.19)$. With this approximation, we may conbine $(3.19)$ and $(3.20)$ and find:
\begin{equation}
\left\{{H}_{g}(\bar{N}_n),....\left\{{H}_{g}(\bar N_3),\left\{{H}_{g}(\bar{N}_2),{H}_{g}(\bar{N}_1)\right\}\right\}...\right\}\big|_{E_{I}^{a}(T_1), A_{b}^{J}(T_1), \bar{N}_i=\mathcal{N}_i}+O(\hbar)+O(d^4)+\epsilon_m=0
\end{equation}
where $\epsilon_m$ denotes generic matter back reactions given by terms involving $\hat{H}_m(N^{\mathcal{N}_i}_{p^*})$. Thus, the emergent gravitational fields satisfy the pure gravitational constraints $ H_{g}(\bar N)=G_{g}(\bar{\Lambda})=M_{g}(\bar V)=0$ up to the corrections $O(\hbar)+O(d^4)+\epsilon_m$.

(2) In canonical general relativity, the unit lapse function ($\bar N(\text x)=1$) leads to a time foliation in which the speed of $\phi^0$  is equal to $v_{\phi^0}$ at the spatial slice $\phi^0(\text x)=T_1$. Therefore, the time foliation using $\phi^0$ as the clock is given by $\bar N(\text x)=1/v_{\phi^0}(\text x)$. With $\mathcal N(X)=1/v_{\phi^0}(X,T_1)$, the symmetry implies:
\begin{equation}
\begin{split}
\langle\Psi|\left(\frac{i}{\hbar}\right)\left[\hat{H}(N^{1/v_{\phi^0}(T_1)}_{p^*}), \hat O(e_{X, \Delta X},T)\right]|\Psi\rangle=0
\end{split}
\end{equation}
When $\hat O(e_{X, \Delta X},T)$ is set to be $\hat{J}(e_{X, \Delta X}, T)_{I}$ and $\hat{h}(e_{X, \Delta X}, T)^{\bar{I}}_{\bar{J}}$, $(3.22)$ gives the clock time dynamics of the emergent gravitational fields:
\begin{equation}
\begin{split}
\frac{d}{dT}\bigg|_{T_{1}}E^{a}_{I}(X,T)\rule{370pt}{0pt}
\\
=\bigg\{E^{a}_{i}(X), \left[H_g(\bar N) +M_g(\bar V)+G_g(\bar\Lambda)\right]\bigg\}\bigg|_{E^a_i=E^a_I(T_1), A_a^i=A_a^I(T_1),\bar N={{\nu}^{-1}_{\phi^0}(T_1)},\bar V^a=\bar V^a({\nu}_{\omega}(T_1)), \bar\Lambda^i=\bar\Lambda^i({\nu}_{\omega}(T_1))}\\ +\epsilon_m+ O(\hbar)+O(d)\rule{362pt}{0pt}
\\\\
\frac{d}{dT}\bigg|_{T_{1}}  A^{J}_{b}(X,T)\rule{370pt}{0pt}
\\
=\bigg\{ A^{j}_{b}(X), \left[H_g(\bar N) +M_g(\bar V)+G_g(\bar\Lambda)\right] \bigg\}\bigg|_{E^a_i=E^a_I(T_1), A_a^i=A_a^I(T_1),\bar N={{\nu}^{-1}_{\phi^0}(T_1)},\bar V^a=\bar V^a({\nu}_{\omega}(T_1)), \bar\Lambda^i=\bar\Lambda^i({\nu}_{\omega}(T_1))}\rule{2pt}{0pt}\\ +\epsilon_m+ O(\hbar)+O(d)\rule{362pt}{0pt}\\
\end{split}
\end{equation}
 Up to the corrections $\epsilon_m+ O(\hbar)+O(d)$, these are general relativity's equations of motion in the Ashtekar formalism, in the gauge $\{\bar N, \bar V, \bar\Lambda\}$ determined by the matter coordinates and frames. Specifically, we have:
\begin{equation}
\bar V^a\big|_{v_{\Phi}=\rule{2pt}{0pt}v_{{}_{f\cdot V_I}}=\rule{2pt}{0pt} v_{{}_{\bar f\cdot\bar V^I}}=\rule{2pt}{0pt}v_{{}_{g\cdot U_{\bar I}}}=\rule{2pt}{0pt}v_{ {}_{\bar g \cdot\bar U^{\bar I}}}=0}=\bar\Lambda^i\big|_{v_{\Phi}=\rule{2pt}{0pt}v_{{}_{f\cdot V_I}}=\rule{2pt}{0pt} v_{{}_{\bar f\cdot\bar V^I}}=\rule{2pt}{0pt}v_{{}_{g\cdot U_{\bar I}}}=\rule{2pt}{0pt}v_{ {}_{\bar g \cdot\bar U^{\bar I}}}=0}=0
\end{equation}
which corresponds to the gauge with the world lines $X=const$ perpendicular to the spatial slice at $T_1$.

We now look into the correction terms in $(3.19)$, $(3.21)$ and $(3.23)$. For this paper we assume that the matter back reactions $\epsilon_m$ are small and focus on the gravitational corrections, which are:
1) The corrections denoted by $O(\hbar)$ resulting from the uncertainty principle. All the fields are quantum mechanical in the model, so there are quantum fluctuations not only in gravitational fields, but also in the matter coordinates and frames. Among the corrections, there are terms coming from regularization of the inverse triad factor $(\det E)^{-1/2}$ appearing in ${H}_g$ by the commutator between the total volume and holonomy operators in $(3.2)$. Thus the quantum Hamiltonian constraint, which is constructed to be always finite, deviates from the classical Hamiltonian constraint when the emergent $E$ fields approaches the scale of $\l_p^2$. Nevertheless, the terms in $O(\hbar)$ are negligible in large $E$ field regions, when the competing classical terms are much greater. 
2) The corrections denoted by $O(d)$ resulting from the discretization of space. Remarkably, these terms are of zeroth order of $\hbar$ and could still dominate in large scales when the quantum effects are insignificant. Recall that $d$ represents the spatial coordinate gap, which is intrinsically finite due to the discretized spatial points $\{p^*_{ n}\}$. The effect of the finite value of $d$ comes in two parts. First, the equations governing the emergent fields are difference equations that approximate the classical differential equations, and therefore it contains corrections of order $O(d)$. Second, the use of holonomies instead of the $A$ fields as configuration variables introduces corrections of higher orders of $|A|d$. In the regularized classical Hamiltonian constraint, the nonlinear terms in the holonomies contribute to errors, which vanish in the limit of the paths shrinking to zero length. In the quantum theory, space is discrete and $d$ is finite, and thus the corrections are intrinsically finite. Suppressed by small $d$ value when $A$ fields are bounded, the holonomy corrections may become important when $A$ fields become singular. Overall, we see that all the corrections denoted by $O(d)$ are negligible when the emergent gravitational fields are nonsingular and vary nicely across a coordinate gap $d$.

When all these corrections are suppressed, general relativity emerges from the semi classical limit of the state $|\Psi\rangle\in \mathbb H$. It should be emphasized that the model shares the kinematics of loop quantum gravity, and the modified Hamiltonian constraint $\hat H( N_p)$ preserves most of the original features. As a result, the key features of loop quantum gravity-- quantum geometry, inverse triad corrections and holonomy corrections--  are all faithfully carried by the model. Moreover, the model provides a set-up to explicitly calculate the correction terms for the dynamics of emergent gravitational fields, including matter back reactions. It is of great interest to see how these corrections behave near the initial singularity, or near a black hole. In  mini or midisuperspace quantum cosmology incorporating the key features of the loop quantum gravity-- loop quantum cosmology-- the quantum and holonomy corrections have been extensively studied. Among the remarkable results of these models is that the initial singularity can be replaced with a well-behaved bouncing of the scale factor, preceded by a contracting universe \cite{lqc1}\cite{lqc2} and followed by a built-in slow-roll inflationary phase \cite{test}\cite{test1}\cite{test2}. These predictions may be testable, and give distinguishable signals in the spectrum of cosmic microwave background radiation \cite{test}\cite{test1}\cite{test2}. It  is then important to ask whether loop quantum cosmology emerges from a certain symmetrical semi-classical limit of loop quantum gravity, and what additional details the full theory would provide regarding the predicted signals. Explicitly evaluating the correction terms in our model could provide answers to these questions. A concrete, affirmative result for the scale-factor bouncing has been obtained in \cite{lin}. Along with many other calculable physical predictions, the model serves as a promising testing ground for ideas from loop quantum gravity.

\section{ Conclusion}

The model proposed here has the same kinematics of loop quantum gravity coupled to matter fields, which is given by knot states. To bypass the difficulties faced by the standard approach, the model uses the modified graph-preserving Hamiltonian constraint operator. This modified constraint operator enables the construction of the model's physical Hilbert space through a group averaging procedure, based on concrete assumptions. Strictly respecting background independence, the model utilizes its matter sector to provide the coordinates and frames to describe the local gravitational observables. The coherent state $|\Psi\rangle$ for the local gravitational observables give rise to classical gravitational fields which obey vacuum general relativity up to the matter back reactions and the quantum gravitational corrections. The corrections have clear interpretations, and the model provides a set-up to explicitly calculate them. It would  be a valuable next step to see how these corrections behave near the initial singularity, and near the singularity of a black hole.

\section{Acknowledgments}
I would like to thank Prof. Steven Carlip for his advice and instruction on conveying this work with minimum distractions from the core ideas. This work was supported in part by Department of Energy grant DE- FG02- 91ER40674.

\end{document}